\documentclass[12pt]{article}
\usepackage{epsfig,amsmath,amssymb,graphics,color,calc} 
\usepackage{cite}

\usepackage{amsmath} 
\usepackage{amsthm}
\usepackage{amssymb}
\usepackage{amsopn}
\newcounter{appendice}

\newtheorem{teo}{Theorem}[section]      \newtheorem{pro}[teo]{Proposition}
\newtheorem{defi}[teo]{Definition}      \newtheorem{lem}[teo]{Lemma}
\newtheorem{cor}[teo]{Corollary}        \newtheorem{rem}[teo]{Remark}
\newtheorem{con}[teo]{Condition}

\newcommand{\bteo}[1]{\begin{teo}\label{#1}}
\newcommand{\bpro}[1]{\begin{pro}\label{#1}}
\newcommand{\bdefi}[1]{\begin{defi}\label{#1}}
\newcommand{\blem}[1]{\begin{lem}\label{#1}}
\newcommand{\bcor}[1]{\begin{cor}\label{#1}}
\newcommand{\brem}[1]{\begin{rem}\label{#1}}
\newcommand{\bcon}[1]{\begin{con}\label{#1}}

\newcommand{\eteo}{\end{teo}}   \newcommand{\epro}{\end{pro}}
\newcommand{\edefi}{\end{defi}} \newcommand{\elem}{\end{lem}}
\newcommand{\ecor}{\end{cor}}   \newcommand{\erem}{\end{rem}}
\newcommand{\econ}{\end{con}}

\renewcommand{\eqref}[1]{(\ref{#1})}

\newcommand{\be}[1]{\begin{equation}\label{#1}}
\newcommand{\bea}[1]{\begin{eqnarray}\label{#1}}
\newcommand{\besn}{\begin{equation*}}
\newcommand{\beasn}{\begin{eqnarray*}}

\newcommand{\su}{\subset}

\newcommand{\diam}{\mathop{\rm diam}\nolimits}
\newcommand{\dis}{\mathop{\rm d}\nolimits}

\newcommand{\ten}{\rightarrow}

\renewcommand{\a}{\alpha}       \renewcommand{\b}{\beta}        
         \renewcommand{\d}{\delta}       
\newcommand{\e}{\varepsilon}    \newcommand{\f}{\varphi} 
\newcommand{\g}{\gamma}         
    
\newcommand{\m}{\mu}

\newcommand{\s}{\sigma}         
\renewcommand{\t}{\tau}         
       
\newcommand{\z}{\zeta}

         \renewcommand{\L}{\Lambda}

    \newcommand{\cL}{\mathcal L}

    \newcommand{\cR}{\mathcal R} 
\newcommand{\cS}{\mathcal S}     
\newcommand{\cU}{\mathcal U}

     \newcommand{\bN}{\mathbb N} 
      
     \newcommand{\bR}{\mathbb R}

     \newcommand{\bZ}{\mathbb Z} 
\newcommand{\newatop}[2]{\genfrac{}{}{0pt}{}{#1}{#2}}

\renewcommand{\tilde}{\widetilde}
\renewcommand{\dis}{\mathrm d}
\renewcommand{\complement}{\mathrm{c}}

\definecolor{light}{gray}{.9}

\begin{document}
\begin{titlepage}
%
%
%
\par\vskip 1cm\vskip 2em

\begin{center}
{\LARGE Gibbsian properties and convergence of the iterates 
\\ 
$\phantom.$ \\
for the Block Averaging Transformation}
\par
\vskip 2.5em \lineskip .5em
{\large
\begin{tabular}[t]{c}
$\mbox{Lorenzo Bertini}^{1} \phantom{m} \mbox{Emilio N.M.\ Cirillo}^{2}
\phantom{m} \mbox{Enzo Olivieri}^{3}$ 
\\
\end{tabular}
\par
}

\medskip
{\small
\begin{tabular}[t]{ll}
{\bf 1} & {\it 
Dipartimento di Matematica, Universit\`a di Roma La Sapienza}\\
&  Piazzale Aldo Moro 2, 00185 Roma, Italy\\
&  E--mail: {\tt bertini@mat.uniroma1.it}\\
\\
{\bf 2} & {\it
Dipartimento Me.\ Mo.\ Mat., Universit\`a di Roma La Sapienza}\\
&  Via A.\ Scarpa 16, 00161 Roma, Italy\\
&  E--mail: {\tt cirillo@dmmm.uniroma1.it}\\
\\
{\bf 3} & {\it
Dipartimento di Matematica, Universit\`a di Roma Tor Vergata}\\
& Via della Ricerca Scientifica, 00133 Roma, Italy\\ 
& E--mail: {\tt olivieri@mat.uniroma2.it}\\
\end{tabular}
}

\bigskip


\end{center}

\vskip 1 em

\centerline{\bf Abstract} 
\smallskip
We analyze the Block Averaging Transformation applied to the 
two--dimensional Ising model in the uniqueness region. We discuss the
Gibbs property of the renormalized measure and the convergence of
renormalized potential under iteration of the map. It turns out that for
any temperature $T$ higher than the critical one $T_c$ the renormalized
measure is strongly Gibbsian, whereas for $T<T_c$ we have only weak
Gibbsianity. Accordingly, we have convergence of the renormalized
potential in a strong sense for $T>T_c$ and in a weak sense for $T<T_c$.
Since we are arbitrarily close to the coexistence region we have a
diverging characteristic length of the system: the correlation length or the
critical length for metastability, or both. Thus, to perturbatively treat
the problem we use a scale--adapted expansion. The more delicate
case is $T<T_c$ where we have a situation similar to that of a disordered
system in the presence of a Griffiths' singularity. In this
case we use a graded cluster expansion whose minimal scale length is diverging
when approaching the coexistence line.

\vskip 0.8 em

\vfill


\bigskip\bigskip
\footnoterule
\vskip 1.0em
{\small 
\noindent
The authors acknowledge the support of Cofinanziamento MURST.
\vskip 1.0em
\noindent
}
\end{titlepage}
\vfill\eject

We want to analyze, from a rigorous point of view, the 
renormalization--group  transformations in statistical mechanics.
We will focus on the well--known Block Averaging Transformation, BAT in the 
sequel, for the standard Ising model. The main questions that we want to 
address concern the Gibbs property of the renormalized measure and the 
convergence of the renormalized potential under iteration of the map.
We shall use perturbative techniques based on cluster expansions.
We mention that, in order to study the Gibbs properties of stochastic fields,
different perturbative approaches have been developed 
(see e.g.\ \cite{[MRSvM],[BKL]}) and also percolation arguments have been 
successfully applied \cite{[vEMSS]}.  

Let us start with the definition of the system,
called {\it object system}, to which the renormalization group map 
is applied. We let $\cS:=\otimes_{x\in\cL}\cS_x$ be the state space, 
with $\cS_x:=\{-1,+1\}$ the single site state space, 
$\cL:=\bZ ^d$ the lattice, and $d\in\bN$ its dimension.
For $\Lambda\subset\cL$ we write $\cS_\Lambda:=\otimes_{x\in\Lambda}\cS_x$.
The Hamiltonian in a finite volume $\Lambda\subset\cL$ with boundary
condition $\tau\in\cS_{\cL\setminus\Lambda}$ is given by 
\begin{equation}
\label{objham}
H^\tau_{\Lambda}(\sigma):=
  -\beta\sum_{\newatop{\{x,y\}\subset\Lambda:}
                      {|x-y|=1}}
   \sigma_x\sigma_y
  -\beta\sum_{\newatop{x\in\Lambda,y\not\in\Lambda:}
                      {|x-y|=1}}
   \sigma_x\tau_y
  -\beta h\sum_{x\in\Lambda}\sigma_x
\end{equation}
for any $\sigma\in\cS_\L$, where 
$\beta={1/T}>0$ is the inverse of the temperature and
$h\in\bR$ the magnetic field.
The corresponding finite volume Gibbs measure is
\begin{equation}
\label{objgib}
\mu^\tau_{\beta,h,\Lambda}(\sigma):= 
  \frac{\exp\{-H^\tau_{\Lambda}(\sigma)\}}
       {\displaystyle\sum _{\eta\in\cS_\L}\exp\{-H^\tau_{\Lambda}(\eta)\}}
\end{equation}
for any $\sigma\in\cS_\L$.
We denote by $\mu=\mu_{\beta,h}$ the unique infinite--volume Gibbs measure
in the region 
\begin{equation}
\label{u}
\cU=\{\b<\b_c\}\cup \{\b>\b_c, \,h
\neq 0\},
\end{equation}
 where
$\beta_c=1/T_c$ is the inverse critical temperature. The region $\cU$ is 
obtained by excluding the critical point from the uniqueness region. 

We define, now, the Block Averaging Transformation. 
Let $\cL^{(\ell)}:=(\ell\bZ)^d$, for $\ell\in\bN$, and partition
$\cL$ as the disjoint union of $\ell$--blocks $Q_\ell(i):=Q_\ell(0)+i$,
where $i\in\cL^{(\ell)}$ and $Q_\ell=Q_\ell(0)$ is the cube of side
$\ell$ with the origin the site with the smallest coordinates.
For $I\subset\subset\cL^{(\ell)}$, with $\subset\subset$ meaning finite
subset of, we set $Q_\ell(I):=\bigcup_{i\in I}Q_\ell(i)$,
We associate with each $i\in \cL^{(\ell)}$ 
a {\it renormalized} or {\it image spin} $m_i$ taking values in
\begin{equation}
\label{rinspa}
\cS^{(\ell)}_i:= 
\bigg\{
 \frac{-\ell^d-\ell^d\bar m}{\sqrt{\ell^d \chi}},\,
 \frac{-\ell^d+2-\ell^d\bar m}{\sqrt{\ell^d \chi}},\,
 \dots,\,
 \frac{\ell^d-\ell^d\bar m}{\sqrt{\ell^d \chi}}
\bigg\}
\end{equation}
where we have denoted by  
$\bar m:=\bar m_{\beta,h}=\m_{\b,h}(\sigma_0)$ the equilibrium
magnetization and by
$\chi:=\chi(\b,h)=\sum_{x\in\cL}
        [\mu_{\b,h}(\s_0 \s_x)-\mu_{\b,h}(\s_0 )\mu_{\b,h} (\s_x)]$ 
the susceptibility.
We next define the renormalized or {\it image} measure 
$\m^{(\ell)}= \m^{(\ell)}_{\b,h}$ on the renormalized or 
image space
$\cS^{(\ell)}:=\otimes_{i\in\cL^{(\ell)}}\cS^{(\ell)}_i$ 
via its finite--dimensional distributions.
Let $I\subset\subset\cL^{(\ell)}$, 
set $\cS^{(\ell)}_I:=\otimes_{i\in I}\cS^{(\ell)}_i$, and pick 
$\tilde m\in\cS^{(\ell)}_I$, then set 
\begin{equation} 
\label{rinmes}
\m^{(\ell)}_{\b,h}\big(\{m\in\cS^{(\ell)}:\,m_I=\tilde m\}\big):=  
 \int_\cS 
  {\rm d}\mu_{\beta,h}(\sigma) 
   \prod_{i\in I} \delta(M_i(\sigma_{Q_\ell(i)})-\tilde m_i)
\end{equation}
where for all $i\in\cL^{(\ell)}$ and $\eta\in\cS_{Q_\ell(i)}$ we have set
\begin{equation}
\label{rinspi}
M_i(\eta):=\frac{1}{\sqrt{\ell^d\chi}}
           \sum_{x\in Q_\ell(i)}[\eta_x-\bar m]
\end{equation}
We write $\mu_{\beta,h}^{(\ell)}=T^{(\ell)}\mu_{\beta,h}$ and note that 
$T^{(\ell)}\,T^{(\ell')}=T^{(\ell\ell')}$. 
The image measure $\m^{(\ell)}_{\b,h}$ represents the distribution of 
the empirical block magnetization $M(\s_{Q_\ell(i)})$, centered and 
normalized, under the object measure $\m_{\b,h}$. 
Various other renormalization group maps, for instance the 
decimation and the majority rule, have been considered in the literature
and successfully applied also to other object systems different
from the standard Ising model, see \cite{[EFS],[NL]}. 

To be concrete in this paper we focus on the standard Ising model. 
However, we stress that the results discussed here  
can be stated in the general setup of lattice spin systems with finite state 
space and finite--range interaction, see \cite{[BCO],[BCOabs],[BCObat]}.
In particular the definitions concerning strong mixing and 
finite--size conditions will be given only for the Ising model.

Our main goal is the study of the map on the potential induced by the 
$T^{(\ell)}$ that has been defined on the infinite--volume measure.
A preliminary condition for this program is that the renormalized 
measure is {\it strongly} or
{\it weakly} Gibbsian  with respect to a
{\it renormalized potential}; where we say that a
stochastic field is strongly resp.\ weakly Gibbsian if its family of 
conditional probabilities has the Gibbsian form with a potential absolutely 
{\it uniformly} resp.\ {\it pointwise almost surely} converging. Thus
in both cases the DLR equations, see (\ref{i:dlr}) below, 
are satisfied but with different strengths in the convergence
properties of the potential. We refer to \cite{[EFS]} for a general 
description of the Gibbs formalism, especially in connection with 
renormalization--group maps, and to \cite{[DS5],[MRSvM]}
for a discussion of the weak Gibbs property.   

We introduce now the finite--volume setup.
Let $I\subset\subset\cL^{(\ell)}$ be a finite box in $\cL^{(\ell)}$ and 
consider the corresponding box  $\Lambda=Q_\ell(I)\subset\cL$. 
We introduce the {\it renormalized Hamiltonian} 
$H^{(\ell),\tau}_I$ with boundary condition 
$\tau\in\cS_{\cL\setminus\Lambda}$ by setting 
\begin{equation}
\label{rinham}
e^{-H^{(\ell),\tau}_I(m)}=
  \sum_{\sigma\in\cS_\Lambda} 
    e^{-H^\t_\L(\sigma)}
    \prod_{i\in I}\delta(M_i(\s_{Q_\ell(i)})-m_i)
\end{equation}
for each $m\in\cS^{(\ell)}_I$.
In the computation of the {\it renormalized potential} associated with the 
renormalized Hamiltonian $H^{(\ell),\tau}_I$,
a crucial role is played by the {\it constrained systems} obtained 
by conditioning the object system to a fixed renormalized spin configuration.
More precisely, the equilibrium probability measure of the constrained 
model associated with the renormalized configuration $m\in\cS^{(\ell)}_I$
on the finite volume $\Lambda=Q_\ell(I)\subset\subset\cL$ is given by
\begin{equation}
\label{conmis}
\m^{(\ell),\tau}_{m,\Lambda}(\sigma):=
  \frac{e^{-H^\tau_\Lambda(\sigma)}
        \prod_{i\in I}\delta(M_i(\s_{Q_\ell(i)})-m_i)}
       {\sum_{\eta\in\cS_\Lambda}e^{-H^\t_\Lambda(\eta)} 
          \prod_{i\in I}\delta(M_i(\eta_{Q_\ell(i)})-m_i)}
\end{equation}
for all $\sigma\in\cS_\Lambda$. 
Notice that from (\ref{rinham}) it follows that the renormalized Hamiltonian 
$H^{(\ell),\tau}_I(m)$ is equal to minus the logarithm of the partition 
function of the corresponding constrained system which is defined as 
\begin{equation}
\label{ham}
Z^{(\ell),\t}_{m,\Lambda}:=
 \sum_{\sigma\in\cS_\Lambda} 
  e^{-H^\tau_\Lambda(\sigma)} 
    \prod_{i\in I}\delta(M_i(\s_{Q_\ell(i)})-m_i)
\end{equation}
In the case of BAT the measure $\m^{(\ell),\t}_{m,I}$ can be called 
{\it multicanonical}, because it is nothing but the original measure 
{\it constrained} to the assigned magnetizations in the
$\ell$--blocks contained in $\Lambda$. 
Of course $\m^{(\ell),\t}_{m,I}$ does not depend at all on the magnetic
field $h$.

Sometimes it happens that the image measure, obtained by 
applying some renormalization group map to a Gibbs object measure 
corresponding, for instance, to a short range object potential, is
not Gibbsian. This pathology often consists of a non--vanishing
dependence of  conditional probabilities of the image measure on
arbitrarily far conditioning  image spins.

It has been shown in \cite{[EFS]} that for any even value of 
$\ell\in\bN$ there exists $\beta_0=\beta_0(\ell)$ (large)
such that the renormalized measure
$\m^{(\ell)}_{\beta,h}$, defined in (\ref{rinmes}),
at any $h$ and any inverse temperature $\beta\ge\beta_0$ is 
non--Gibbsian  as a consequence of violation of {\it quasi--locality}, 
a continuity property of its conditional probabilities 
which constitutes a necessary condition for Gibbsianity.
This kind of influence from infinity is, in turn, a consequence of a first order
phase transition with long range order of a particular 
constrained model namely, the one corresponding to  
$m_i=0$ for all $i\in\cL^{(\ell)}$. Thus the
existence of just one bad image configuration, giving rise to a bad  
constrained system, is sufficient to induce the violation of the 
Gibbs property of the image measure. On the other hand it is
reasonable to expect, and proven in \cite{[HK]} in a specific context,
that when {\it all} the constrained systems are 
well--behaved the image measure is Gibbsian. As we shall see in the sequel,
this will be the case for the two--dimensional Ising Model above $T_c$.
It is clear that the above described pathology for BAT applied to 
the low--temperature Ising model is completely independent of the value 
of the magnetic field $h$ acting on the object system. 
On the other hand it is also clear that this ``bad"
configuration inducing non--Gibbsianity, is very atypical with respect to
$\m^{(\ell)}_{\beta,h}$ for  $h\neq 0$.
It is thus reasonable to expect at least the validity of a weaker property 
of Gibbsianity.
For $h=0$ we can consider the BAT for the extremal measures $\mu_\pm$;
we expect weak Gibbsianity also in this case. 

\bigskip
We want to present now  results on strong Gibbsianity above $T_c$ and  on weak
Gibbsianity below $T_c$ together with results on convergence of the 
iterated renormalized potential. 
All these results are based on a suitable strong mixing condition  
for the object system namely, exponential decay of finite--volume 
truncated expectations with a rate independent of the boundary conditions and  
of the volume chosen from a suitable class. 
More explicitly, given an integer $\ell_0$, 
we say that the measure $\mu^\tau_{\beta,h,\Lambda}$ satisfies 
SM($\ell_0$) if there exist two constants $C,\g>0$ such that for
every pair of local functions $f,g$ with supports 
$S_f,S_g$, and every volume of the form 
$\Lambda=Q_{\ell_0}(I)$, with $I\subset\subset\cL^{(\ell_0)}$, containing 
$S_f,S_g$, we have that 
\begin{equation}
\label{sm}
\sup _{\t\in \cS_{\cL\setminus \L}} |\mu^{\t}_{\b,h,\L}(f;g)|
\le C (|S_f|\wedge|S_g|)\,  \|f\|_{\infty}\, \|g\|_{\infty} \, 
    e^{-\g\,\dis(S_f,S_g)} 
\end{equation}
where for $\Delta,\Delta'\subset\cL$ we have set 
$\dis(\Delta,\Delta'):=\inf\{|x-x'|,\,x\in\Delta,x'\in\Delta'\}$. 

To simplify the exposition we will mainly consider the two--dimensional case;
the higher--dimensional case will be briefly discussed.
In two dimensions it has been proved, see \cite{[MOS],[ScS]},
that condition SM($\ell_0$) is satisfied in the whole uniqueness region 
$\cU$ for some $\ell_0$ depending on $\b,h$.
We observe that there exist two characteristic lengths of the Ising model
namely, the correlation length and the {\it critical length for metastability},
which diverge when the closure of the coexistence line 
$\cU^\complement:=\{\beta\ge\beta_c,h=0\}$ is approached. The first diverges 
in the limit $h=0$ and $\beta\to\beta_c^-$, the latter diverges 
as $1/h$ in the limit $\beta>\beta_c$ fixed and $h\to0$. 
The critical length for metastability 
represents the minimal size of a droplet whose growth is energetically 
favorable and, at the same time, the minimal length required to 
{\it screen} the effect of a boundary condition opposite to the field.
It is clear that a perturbative description outside $\cU^\complement$, 
uniform in the boundary conditions, has to involve a characteristic 
length depending on $\beta,h$ and diverging when the distance from
$\cU^\complement$ tends to zero. 
Notice that the critical length for metastability and 
the correlation length can even diverge simultaneously.  

To extract the renormalized potentials from the renormalized
Hamiltonian $H^{(\ell),\tau}_I(m)$, where $m\in\cS^{(\ell)}_I$,
a possible strategy is to use a perturbative expansion for the 
constrained system corresponding to $m$ with a procedure making sense in the
thermodynamic limit. This should work for every $m\in\cS^{(\ell)}_I$.
For instance when the object system is far away from the coexistence line 
then the usual high temperature or high magnetic field expansions for the 
constrained models are sufficient to compute the renormalized potentials,
see \cite{[GP],[I],[Cam],[Ka]}. However, in order to
get close to the coexistence line, we certainly have to use other, more
powerful, perturbative theories. 

We discuss now these different perturbative
theories in the concrete case of systems above their critical 
temperature $T_c$.
Usual high temperature expansions work only for temperatures $T$ sufficiently
larger than $T_c$; they basically involve perturbations around a universal
reference system composed of independent spins; in other words  
the small parameter is the inverse temperature and all interactions are
expanded treating, in this way, every lattice system in the same manner. 
In \cite{[O],[OP]} another perturbative expansion has been 
introduced, around a non--trivial model--dependent reference system,
that we   call {\it scale--adapted expansion}. The small parameter is
no more
$\b=1/T$ but, rather, the ratio between the correlation length 
(at the given temperature $T>T_c$) and the length scale $L$ at which we 
analyze our system. The geometrical objects (polymers) involved in  
the scale--adapted expansion live on the scale $L$ whereas in the usual
high and low temperature or high magnetic field 
expansions they live on scale one. 
Of course the smaller is $T-T_c$ the larger has to be taken the length $L$.
A similar situation occurs for low temperature Ising ferromagnets with 
generic boundary conditions at arbitrarily small  
but non zero magnetic field $h$ with a diverging critical
length  of order $1/h$.
Also in this case we 
have to look at our system on a scale sufficiently larger than 
the critical length; note that at low temperature, far from the 
critical point, the correlation length is of order one. 

The scale--adapted expansions are based on a suitable {\it finite size
condition} saying, roughly speaking, that if we look at the Gibbs 
measure in a box of sufficiently large side length $L$, then, uniformly in 
the boundary conditions, the corresponding truncated correlations decay 
sufficiently fast in terms of $L$. A possible formulation is the following,
see \cite{[O],[OP],[MO2]},
we say that condition C($\e,L$) is satisfied if there exist $\e>0$ 
and $L\in\bN$ such that
\begin{equation}
\label{fsc}
\sup_{\tau\in\cS_{\cL\setminus Q_L}}\,
\sup_{\newatop{x,y\in Q_L:}
              {|x-y|\geq L-1}}
|\m^\t_{\b, h,Q_L}(\s_x;\s_y)| < \frac{\e}{L^{2(d-1)}}
\end{equation}
where we recall $Q_L=Q_L(0)$.
It has been proven (see \cite{[O],[OP],[MO2]}) that there
exists a real $\e_0=\varepsilon_0(d)$ --- recall 
$d$ is the dimension of the lattice --- such that if there exists 
$L\in\bN$ with the property that condition C($\e,L$) is satisfied with 
$\e<\e_0$, then strong mixing 
SM($\ell_0$) holds for every $\ell_0$ multiple of $L$.
Conversely, it is immediately seen that if we assume the validity of 
SM($\ell_0$) for some $\ell_0$, then condition C($\e,L$) holds for every 
$\e>0$ and a sufficiently large $L$ multiple of $\ell_0$.
We remark that to get the above result it is not necessary to use the 
cluster expansion (see \cite{[M]}), however this theory is needed if we 
want to prove complete results like analyticity properties of 
thermodynamic and correlation functions.

The basic idea to develop a perturbative theory on the basis 
of condition C($\e,L$) is inspired by the renormalization group 
theory itself and consists in applying a block decimation procedure. 
Consider a box in $\cL$ of the form $\L=Q_L(I)$, with $I$ a cube in
$\cL^{(L)}$; we identify the blocks $Q_L(i)$ in $\L$ with the points 
$i$ in $I$. We look at the rescaled system as a lattice spin system on 
$\cL^{(L)}$ whose spin variable associated to the site $i\in I$ is 
identified with the block configuration in $\cS_{Q_L(i)}$.
Then we partition $\cL^{(L)}$ as the disjoint union of $2^d$ 
sublattices of spacing $2L$ and enumerate these sublattices with a given 
order. To compute the partition function $Z_\Lambda$ of the 
model, assuming for simplicity periodic boundary 
condition, we start summing over the variables in the first
sublattice, keeping fixed the  variables on the other sublattices. 
Choosing $L$ larger than the range of the interaction, this first sum 
factorizes. Condition C($\e,L$), for small $\e$, implies that opposite 
faces of the cube $Q_L(i)$ are weakly correlated so that after the summation 
over the variables on the first sublattice, we get that the effective 
interaction between the surviving variables, sitting on the other
sublattices, is weak.
Then we iterate this procedure by successively summing over the $L$--blocks 
variables sitting on subsequent sublattices.
In this way we end up with an expression of the partition function 
of the form
\begin{equation}
\label{fatt}
Z_\Lambda =\bar Z_\L\, \Xi_\L
\end{equation}
with $\bar Z_\Lambda$ given as the product of partition functions on 
suitable domains with sizes of order $L$ and $\Xi_\L$ the partition 
function of a gas of polymers given as suitable unions of $L$--blocks
whose only interaction is a hard core exclusion:
\begin{equation}
\label{paly}
\Xi_\L = 1 + \sum_{n\ge 1}
     \sum_{\newatop{R_1,\dots , R_n\in \cR_\L :}
                   {R_i \cap R_j =\emptyset}}
     \z(R_1)\cdots\z(R_n)
\end{equation}
where $\cR_\L$ is the set of polymers contained in $\L$ and 
$\z:\cR_\L\to \bR$ is the activity.
Two overlapping polymers are called {\it incompatible}.
The validity of condition C($\e,L$) with small $\e$ implies 
that the polymer system is in the small
activity regime. More precisely the following estimate holds
\begin{equation}
\label{sti}
\sum_{R\ni O} |\z(R)| \leq \d(\e)\;\;\;
 {\rm with}\;\;\;\d(\e)\to 0\;\;\;{\rm as}\;\;\;\e \to 0
\end{equation}
where $O=(0,\dots,0)$ is the origin of $\cL$. 

Suppose that we want to treat perturbatively the free energy given 
as $-(1/|\L|)\log Z_\L$.
Taking the logarithm of $Z_\L$ by \ref{fatt} we get two terms. 
The logarithm of $\bar Z_\L$
gives rise to a sum of local terms whereas the logarithm 
of $\Xi_\L$ can be treated by means
of the general theory of cluster expansions (see for instance 
\cite{[GMM],[KP]}) which is based
on a condition like (\ref{sti}) with a sufficiently small $\d$. 
We get an expansion of the form
\begin {equation}
\label{ce}
\log \Xi_\L = \sum _{R_1 , \dots R_n \in \cR_\L}\f_T (R_1 , \dots R_n) 
 \z(R_1)\cdots\z(R_n) 
\end {equation}
where the combinatorial factor 
$\f_T(R_1,\dots,R_n)$ vanishes when $R_1,\dots,R_n$ do not
form a globally incompatible set.
We observe that the building bricks of the above described scale--adapted 
cluster expansion lives on scale $L$; we look at our system of 
$L$--block variables and we never go below this minimal scale.

Before going back to the discussion of BAT we want  
to mention here that there is another,
stronger notion of strong mixing, originally introduced
for general short--range lattice systems 
by Dobrushin and Shlosman in \cite{[DS1],[DS2]}, before the 
papers \cite{[O],[OP]} appeared. It has
been called by the authors {\it Complete Analyticity} (CA): it requires 
exponential decay of truncated correlations for {\it all} finite or 
infinite domains $\L$ (of arbitrary shape).
This point of view contrasts the one of scale--adapted expansions.
Indeed in Dobrushin--Shlosman's approach there is no minimal scale 
length and there are
examples where SM($\ell_0$) holds for $\ell_0$ sufficiently large but 
Dobrushin--Shlosman's
Complete Analyticity fails for domains with anomalous ratio surface/volume.

Scale--adapted perturbative
theory gives rise to a notion that can be called {\it Restricted
Complete Analyticity} (RCA) or {\it Complete Analyticity for Regular
Domains}; here ``regular" means ``multiple" of a sufficiently large box.
Dobrushin and Shlosman also developed a finite size condition involving 
{\it all} the possible subsets of a given, sufficiently large box and not 
just the box itself like \cite{[MO2]}: in some respects 
Dobrushin--Shlosman's approach can be still seen as a perturbative theory on 
scale one. 
For the two--dimensional standard Ising model, as we said before, 
RCA has been proven in the whole region $\cU$ \cite{[ScS]}.
On the other hand CA has been conjectured to hold for any 
$T>T_c$; however a proof is still missing. For $d\ge3$ even RCA is
not expected to hold at low--temperature for suitable nonvanishing magnetic 
field as a consequence of a layering phase transition \cite{[Ba],[CM]}.

The Gibbs properties of the renormalized measure are particularly relevant,
from the physical point of view, in a neighborhood of the critical point.
 Actually it
is possible that even if the object system is critical it may happen that
the constrained systems are in the one phase, weakly coupled regime, so that
the renormalized potential is still well defined,
see \cite{[HK],[BMO],[CiO]}.

On the other hand   examples of non Gibbsianity are given in \cite{[E1]},
\cite{[E2]}.

Let us now state the main results of \cite{[BCO]} on strong Gibbsianity and 
convergence above $T_c$ in two dimensions.

\bteo{hight}{\normalfont\normalsize\textrm(Bertini, Cirillo, Olivieri 1999)}
Consider a two--dimensional
Ising system with $\b <\b_c$ and $h \in \bR$ given.
Then there exists $\ell_0\in\bN$ such that for any
$\ell$ large enough multiple of $\ell_0\;$ $ \m^{(\ell)}_{\b,h}$ is Gibbsian
in the sense that for each $Y\subset\subset\cL^{(\ell)}$
and for each local function $f:\cS_Y^{(\ell)}\to\bR$ we have
\begin{equation}
\label{i:dlr}
\begin{array}{l}
{\displaystyle
\mu^{(\ell)}(f)=
 \int_{\cS^{(\ell)}}
  \mu^{(\ell)}(dm')\, \frac{1}{Z_Y(m')}\,
  \sum_{m\in\cS_Y^{(\ell)}}f(m)}\\
{\displaystyle
\;\;\;\;\;\;\;\;\;\;\;\;\;\;\;\;\;\;\;\;\;\;\;\;\;\;
\times\exp\Big\{\sum_{X\cap Y\neq\emptyset}
               [\psi_X^{(\ell)}(m_{Y \cap X }m'_{Y^\complement\cap X})+
               \phi_X^{(\ell)}(m_{Y \cap X }m'_{Y^\complement \cap X})]\Big\}}
\end{array}
\end{equation}
where
\begin{equation}
\label{i:fpart}
 Z_Y(m') =\sum_{m\in\cS_Y^{(\ell)}}
       \exp\Big\{\sum_{X\cap Y\neq\emptyset}
       [\psi_X(m_{Y \cap X }m'_{Y^\complement\cap X})+\phi_X(m_{Y \cap X
}m'_{Y^\complement\cap X})]\Big\}
\end{equation}
and the family
$\{\phi^{(\ell)}_X+\psi^{(\ell)}_X,\; X \su \su \cL^{(\ell)} \},$
$\phi^{(\ell)}_X,\psi^{(\ell)}_X: \cS^{(\ell)}_X \to \bR$
is translationally invariant and
satisfies the uniform bound
\begin{equation}
\label{i:finnorm}
\sum_{X \ni 0}   e^{\a |X|} \sup_{m_X \in \cS^{(\ell)}_X}  \left (
\left|\psi^{(\ell)}_X(m_X)\right|+\left|\phi^{(\ell)}_X (m_X)   \right| \right )
 < \infty.
\end{equation}
for a suitable $\a>0$.
Moreover
$\exists\kappa\in\bN$: $\psi_{X}^{(\ell)}= 0$ if
$\diam(X)\ge\kappa$.
Finally we have that for the same $\alpha$ as in \eqref{i:finnorm}
\begin{equation}
\label{i:conver}
\lim_{\ell\ten\infty}
\sum_{X \ni 0}  e^{\a |X|} \sup_{m_X \in \cS^{(\ell)}_X}
\left|\phi^{(\ell)}_X (m_X)   \right|
 =0,
\end{equation}
$\psi^{(\ell)}_{\{i\}}(m_i)=m_i^2/2$ for $i\in\cL^{(\ell)}$
and there exists $a>0$ such that
$$
\lim_{\ell\ten\infty} 
\sup_{\newatop{m_X\in \cS^{(\ell)}_X}
              {|m_i|\le\ell^a, i\in X}
     }
\left| \psi^{(\ell)}_X (m_X)  \right| = 0
\quad \quad {\rm for }~~ |X|\ge 2
$$
\eteo
 
Notice that, by the composition rule $T^{(\ell)}T^{(\ell')} = T^{(\ell
\ell')}$, taking the limit 
$\ell\to\infty$ is equivalent to indefinitely iterating the 
map $T^{(\ell_0)}$. 

To prove the above theorem we use, for the partition function of the 
constrained system (\ref{ham}), the validity of an expression like 
(\ref{fatt}) with $\Xi_\Lambda$ in the form  
(\ref{paly}) and $\zeta(R)$ satisfying (\ref{sti}). 
This is obtained on the basis of the validity of the finite--size condition 
(\ref{fsc}) with $\varepsilon=\e(\ell)\to0$ as $\ell\to\infty$
{\it uniformly} in the renormalized configuration $m\in\cS^{(\ell)}$.
By using a delicate comparison between multi--canonical and 
multi--grandcanonical ensembles \cite{[BCO]} we see that 
the crucial point to obtain the above uniform finite size condition is 
the validity of the strong mixing condition SM$(\ell_0)$
for the object system {\it uniformly} in the magnetic field $h$. 
This is sufficient only in the two--dimensional case. In higher 
dimension   instead of this one  needs uniformity w.r.t.\ variable magnetic 
field constant in each cube of side $\ell_0$. Uniformity of 
SM$(\ell_0)$ w.r.t\ $h$ fails for any $d\ge2$ below $T_c$, because of the phase 
transition at $h=0$. By only assuming strong mixing of the object
system, without uniformity in $h$, we can expect only weak Gibbsianity since, 
as we said before, for $T<T_c$ violation of strong Gibbsianity  
 is proven in \cite{[EFS]}. 
Let us now state our main results on weak Gibbsianity and convergence of the 
iterates of BAT.

\bteo{t:wgib}
Consider a two--dimensional
Ising model and recall $\cU$ is defined in (\ref{u}). Given 
$(\b,h)\in\cU$,
there exists $\ell_0$ such that for any large enough
$\ell$ multiple of $\ell_0$, $\m^{(\ell)}$ is weakly Gibbsian in the sense 
that it satisfies the DLR equations \eqref{i:dlr} with respect to a potential
$\{\psi_X^{(\ell)}+\phi_X^{(\ell)},\,X\subset\subset\cL^{(\ell)}\}$,
  $\psi_X^{(\ell)},\phi_X^{(\ell)}:\, \cS^{(\ell)}_X\mapsto\bR$,
satisfying the following.

There exists a measurable set $\bar\cS^{(\ell)}\subset\cS^{(\ell)}$, such that
$\mu^{(\ell)}(\bar\cS^{(\ell)})=1$, and functions
$r^{(\ell)}_i:\bar\cS^{(\ell)}\mapsto\bN\setminus\{0\}$,
for all $i\in\cL^{(\ell)}$, such that
for each $m\in\bar\cS^{(\ell)}$ if
$X\ni i$ and $\diam(X)\ge r^{(\ell)}_i(m)$ then $\psi_X^{(\ell)}(m)=0$.
Furthermore, for each $i\in\cL^{(\ell)}$ and $m\in\bar\cS^{(\ell)}$
there exists a real $c_i^{(\ell)}(m)\in[0,\infty)$ such that
\begin{equation}
\label{in:mtb5}
\sum_{X\ni i}|\psi_X^{(\ell)}(m_X)|\le c_i^{(\ell)}(m)
\end{equation}
There exists $C$ independent of $\ell$ such that
\begin{equation}
\label{i:eee}
\sup_{i\in\cL^{(\ell)}}
\sum_{X\ni i}
\sup_{m_X\in\cS^{(\ell)}_X}
    |\phi_X^{(\ell)}(m_X)|<C
\end{equation}
For each $i\in\cL^{(\ell)}$ we have $\psi_{\{i\}}^{(\ell)}(m)=m_i^2/2$ and
for each $q\in[1,+\infty)$
\begin{equation}
\label{i:ee2}
\lim_{\ell\to\infty}\,
\sup_{i\in\cL^{(\ell)}}
  \mu^{(\ell)}\Big(\Big|
              \sum_{X\ni i:\,|X|\ge2}\psi_X^{(\ell)}
              \Big|^q\Big)
=0
\end{equation}
and
\begin{equation}
\label{i:ee1}
\lim_{\ell\to\infty}
\sup_{i\in\cL^{(\ell)}}
\sum_{X\ni i} 
\sup_{m_X\in\cS^{(\ell)}_X}
|\phi_X^{(\ell)}(m_X)|
  =0
\end{equation}
\end{teo}

The above theorem is a consequence of a general result saying that in 
order to get weak Gibbsianity for a renormalized potential and convergence of 
the iterates in the above sense, we have only to check the validity, for the 
object system with a given value of $h$, of SM($\ell_0$) for some
$\ell_0\in\bN$,  see \cite{[BCObat]}; 
this, as we said before, is a general result in the region $\cU$ for d=2.
Thus the result of Theorem~\ref{t:wgib} immediately
extends to the case $d>2$, $h\neq0$, and $\beta>\beta_0(d,|h|)$ for a suitable 
function $\beta_0:\bN\times\bR^+\to\bR^+$. 
Indeed in this case strong mixing holds true, see \cite{[MO2]}. 

\bigskip
Let us now discuss the result of Theorem~\ref{t:wgib}; 
in particular we want to analyze the
difficulties in computing the renormalized potential and the 
way to solve them in the
concrete case of a low--temperature Ising system at $h \neq0$.
We know that in the bad regions with $m_i=0$ the corresponding 
constrained system almost undergoes a phase transition with long 
range order so that the conditional probabilities of the renormalized 
measure in these regions are very sensitive to the values of arbitrarily 
far conditioning spins; in other words the corresponding renormalized 
potential does not decay fast enough at large distances.

For $h\neq0$ this is a very atypical configuration; however with small but 
positive probability we have arbitrarily large bad regions with $m_i =0$.
In order to prove weak Gibbsianity the key property is that bad regions are 
far apart: larger
and larger bad regions are sparser and sparser. 
This situation is
similar to that of disordered systems in the presence of a Griffiths' 
singularity \cite{[BCOabs]}. A multi--scale analysis is needed. 

The natural approach, quite complicated from the technical
point of view, is to use a graded cluster expansion.
For disordered systems there are clever methods,
see \cite{[Berretti],[vDKP]},
avoiding cluster expansion, that enable to prove partial results 
like exponential clustering with deterministic rate and random prefactor.
In the case of BAT, in order to compute renormalized potentials in 
the weakly Gibbsian case,
the use of the full theory of graded cluster expansion 
(like the one in \cite{[FI]}) appears to be unavoidable.
Since we want to study a region of parameters arbitrarily 
close to  $\cU^\complement=\{\beta\ge\beta_c,h\neq0\}$, the closure
of the coexistence line, the distinctive character of our graded 
cluster expansion is that the minimal scale may be chosen arbitrarily 
large and diverging as $T\to T_c$ and/or $h\to 0$.
The minimal scale involved in our discussion being divergent, we need to 
use a scale--adapted cluster expansion,
see \cite{[O],[OP],[BCOabs]}, based on a finite--size mixing condition. 

As we observed before in this case, 
contrary to low and high temperature expansion or high magnetic 
field expansion, the small parameter is the ratio between the diverging 
length and the suitably large finite size where the mixing condition holds.
We want to stress again that in our approach, according to the general 
renormalization group ideology, we first fix the values of the 
thermodynamic parameters of the object system and, subsequently, the value of 
the scale of BAT. In other words we {\it take advantage} from choosing the 
scale $\ell$ of the transformation large enough. On the other side we cannot 
exclude that, for given values of $\beta$ and $h$, if
$\ell$ is not sufficiently large weak Gibbsianity ceases to be valid.
In \cite{[MRSvM],[BKL]} the authors study decimation transformation,
see \cite{[EFS]}, at large $\beta$ and {\it arbitrary h}. 
They first fix the scale of the transformation and, subsequently, they 
choose the temperature below which they get weak Gibbsianity.

It seems clear that to get results of convergence of renormalized
potentials when iterating BAT, one has to use a perturbative theory 
based on scale--adapted cluster expansion. 
Even far away from the critical point, in order to get convergence, one
needs to take advantage from choosing larger and larger $\ell$.

In \cite{[Cam]} the author proves convergence results at high temperature 
by making use of a general result \cite{[Hu]}, 
according to which, to get convergence in a suitable sense one needs 
only to prove uniform boundedness in a suitable norm.
In \cite{[Cam]} the author uses a high temperature expansion giving 
rise to a polymer system whose activity is small  
uniformly in $\ell$ for $\ell$ large enough.
This appears to contradict the necessity to use a scale--adapted cluster
expansion, but we want to stress that 
he does not directly prove the convergence 
of the renormalized potential but only uniform boundedness.
This situation is similar to the one of \cite{[S1]} where the author 
uses a low--temperature expansion that converges uniformly in $\ell$ 
but does not deal with the problem of convergence.

We remark, finally, that also for a disordered lattice spin systems 
in the Griffiths' phase and close to criticality, it seems necessary to 
use a graded cluster expansion whose minimal length scale is not one as 
in \cite{[FI]} but diverges as $T\to T_c$, see \cite{[BCOran]}.
We conclude by briefly discussing an example of small random perturbation 
of a strong mixing system possibly close to criticality: a ferromagnetic 
two dimensional Ising system with zero magnetic field and coupling constants
given by i.i.d.\ random variables for different bonds with distribution 
\begin{equation}
\label{rp}
J=\left\{
\begin{array}{ll}
1 & \textrm{with probability } 1-p\\
J_0 & \textrm{with probability } p\\
\end{array}
\right.
\end{equation}
Let $T_c(1)$ be the critical temperature corresponding to a coupling 
constant equal to one. We expect the validity of the following result: 
for all $T>T_c(1)$ there exists $p_0>0$
such that for all $p<p_0$ and $J_0\leq+\infty$ we have, for almost all the
realizations of the disorder, a convergent  multi--scale cluster expansion
whose minimal scale diverges as $T\to T_c(1)$.


 
 
 

\begin{thebibliography}{99}

\bibitem{[Ba]}
A.G.\ Basuev,
``Hamiltonian of the phase separation border and phase transition
of the first kind. I."
{\sl Theor.\ Math.\ Phys.} {\bf 64}, 716--734 (1985).

\bibitem{[BMO]}
G.\ Benfatto, E.\ Marinari, E.\ Olivieri,
``Some numerical results on the block spin transformation
for the 2D Ising model at the critical point."
{\sl J.\ Statist.\ Phys.} {\bf 78}, 731--757 (1995).

\bibitem{[BCO]}
L.\ Bertini, E.N.M.\ Cirillo, E.\ Olivieri,
``Renormalization Group Transformations under strong mixing conditions:
Gibbsianness and convergence of renormalized interactions."
{\sl J.\ Statist.\ Phys.} {\bf 97}, 831--915 (1999).


\bibitem{[BCOran]}
L.\ Bertini, E.N.M.\ Cirillo, E.\ Olivieri,
``Randomly perturbed strong mixing systems: beating
Griffiths' singularity above the critical temperature."
In preparation.

\bibitem{[BCOabs]} 
L.\ Bertini, E.N.M.\ Cirillo, E.\ Olivieri, 
``Graded cluster expansion for lattice systems."
In preparation.

\bibitem{[BCObat]} 
L.\ Bertini, E.N.M.\ Cirillo, E.\ Olivieri, 
``Renormalization Group in the uniqueness region: weak Gibbsianity and 
convergence."
In preparation.

\bibitem{[Berretti]}
A.\ Berretti,
``Some properties of random Ising models."
{\sl J.\ Statist.\ Phys.} {\bf 38}, 483--496 (1985).


\bibitem{[BKL]}
J.\ Bricmont, A.\ Kupiainen, R.\ Lefevere,
``Renormalization group pathologies and the definition of
Gibbs states."
{\sl Comm.\ Math.\ Phys.} {\bf 194}, 359--388 (1998).

\bibitem{[Cam]}
C.\ Cammarota,
``The Large Block Spin Interaction."
{\sl Nuovo Cimento} {\bf B(11) 96}, 1--16 (1986).







\bibitem{[CM]}
F.\ Cesi, F.\ Martinelli,
``On the layering transition of an SOS surface interacting with a wall. 
I. Equilibrium results."
{\sl J.\ Statist.\ Phys.} {\bf 82}, 823--913 (1996).

\bibitem{[CiO]}
E.N.M.\ Cirillo, E.\ Olivieri,
``Renormalization group at criticality and complete analyticity
of constrained models: a numerical study."
{\sl J.\ Statist.\ Phys.} {\bf 86}, 1117--1151 (1997).





\bibitem{[vDKP]}
H.\ von Dreifus, A.\ Klein, J.F.\ Perez,
``Taming Griffiths' singularities: infinite differentiability
of quenched correlation functions.
{\sl Comm.\ Math.\ Phys.} {\bf 170}, 21--39 (1995).




\bibitem{[DS1]}
R.L.\ Dobrushin, S.B.\ Shlosman,
``Constructive Criterion for the Uniqueness of Gibbs Fields."
{\sl Statist.\ Phys.\ and Dyn.\ Syst.}, Birkhauser, 347--370 (1985).

\bibitem{[DS2]}
R.L.\ Dobrushin, S.B.\ Shlosman,
``Completely Analytical Gibbs Fields."
{\sl Statist.\ Phys.\ and Dyn.\ Syst.}, Birkhauser, 371--403 (1985).



\bibitem{[DS5]}
R.L.\ Dobrushin, S.B.\ Shlosman,
``Non-Gibbsian states and their Gibbs description."
{\sl Comm.\ Math.\ Phys.} {\bf 200}, 125--179 (1999).


\bibitem{[E1]}
A.C.D.\ van Enter,
``Ill--defined block--spin transformations at arbitrarily
high temperatures."
{\sl J.\ Statist.\ Phys.} {\bf 83}, 761--765 (1996).

\bibitem{[E2]}
A.C.D.\ van Enter,
``On the possible failure of the Gibbs property for
measures on lattice systems. Disordered systems and statistical
physics: rigorous results."
{\sl Markov Process.\ Related Fields} {\bf 2}, 209--224 (1996).


\bibitem{[EFS]}
A.C.D.\ van Enter, R.\ Fern{\'a}ndez, A.D.\ Sokal,
``Regularity Properties and Pathologies of Position--Space
Renormalization--Group Transformations: Scope and Limitations of
Gibbsian Theory."
{\sl J.\ Statist.\ Phys.} {\bf 72}, 879--1167 (1994).

\bibitem{[vEMSS]}
A.C.D.\ van Enter,  C. Maes, R.H.\ Schonmann, S.B.\ Shlosman,
``The Griffiths singularity random field."
On Dobrushin's way. From probability theory to statistical physics, 51--58, 
Amer.\ Math.\ Soc.\ Transl.\ Ser.\ 2, {\bf 198}, 
Amer.\ Math.\ Soc., Providence, RI, 2000. 




\bibitem{[FI]}
J. Fr\"ohlich, J.Z.\ Imbrie,
``Improved perturbation expansion for disordered systems:
beating Griffiths' singularities."
{\sl Comm.\ Math.\ Phys.}  {\bf 96}, 145--180 (1984).




\bibitem{[GMM]}
Gallavotti G., Martin L\"of A., Miracle Sole S.,  in Battelle Seattle
(1971) Rencontres, A. Lenard, ed.
(Lecture Notes in Physics, Vol. 20, Springer,
Berlin, 1973), pp.162-204.




\bibitem{[GP]}
R.B.\ Griffiths, P.A.\ Pearce,
``Mathematical Properties of Position--Space
Renormalization Group Transformations."
{\sl J.\ Statist.\ Phys.} {\bf 20}, 499--545 (1979).

\bibitem{[HK]}
K.\ Haller, T.\ Kennedy,
``Absence of renormalization group pathologies near the critical
temperature. Two examples."
{\sl J.\ Statist.\ Phys.} {\bf 85}, 607--637 (1996).


\bibitem{[Hu]}
N.M.\ Hugenholtz, 
``On the inverse problem in statistical mechanics."
{\sl  Comm.\ Math.\ Phys.} {\bf 85}, 27--38 (1982).


\bibitem{[I]}
R.B.\ Israel,
``Banach Algebras and Kadanoff Transformations
in Random Fields."
J.\ Fritz, J.L.\ Lebowitz and D.\ Szasz editors
(Esztergom 1979), Vol. II, 593--608 (North--Holland, Amsterdam 1981).

\bibitem{[Ka]}
I.A.\ Kashapov,
``Justification of the renormalization group method."
{\sl Theor.\ Math.\ Phys.} {\bf 42}, 184--186 (1980).

\bibitem{[KP]}
R.\ Koteck\'y, D.\ Preiss,
``Cluster expansion for abstract polymer models."
{\sl Comm.\ Math.\ Phys.} {\bf 103}, 491--498 (1986).







\bibitem{[M]}
F.\ Martinelli,
``An elementary approach to finite size conditions for the exponential 
decay of covariance in lattice spin models." 
{\it On Dobrushin's way. From probability theory to statistical 
physics}, 169--181, Amer.\ Math.\ Soc.\ Trans.\ Ser.\ 2, 198,
{\it Amer.\ Math.\ Soc., Providence, RI.} (2000)



\bibitem{[MO2]}
F.\ Martinelli, E.\ Olivieri,
``Approach to Equilibrium of Glauber Dynamics
in the One Phase Region I. The Attractive Case."
{\sl Commun.\ Math.\ Phys.} {\bf 161}, 447--486 (1994).




\bibitem{[MOS]}
F.\ Martinelli, E.\ Olivieri, R.\ Schonmann,
``For $2$--$D$ Lattice Spin Systems Weak Mixing Implies Strong Mixing."
{\sl Commun.\ Math.\ Phys.} {\bf 165}, 33--47 (1994).


\bibitem{[MRSvM]}
C.\ Maes, F.\ Redig, S.\ Shlosman, A.\ Van Moffaert,
``Percolation, path large deviations and weakly Gibbs states."
{\sl Commun.\ Math.\ Phys.} {\bf 209}, 517--545 (2000).




\bibitem{[NL]}
Th.\ Niemeijer,  M.J.\ van Leeuwen,
``Renormalization theory for Ising--like spin systems."
In ``Phase Transitions and Critical Phenomena",
vol. 6, Eds. C. Domb, M. S. Green  (Academic Press, 1976).

\bibitem{[O]}
E.\ Olivieri,
``On a cluster expansion for Lattice Spin Systems: a Finite
Size Condition for the Convergence."
{\sl J.\ Statist.\ Phys.} {\bf 50}, 1179--1200 (1988).

\bibitem{[OP]}
E.\ Olivieri, P.\ Picco,
``Cluster expansion for $D$--Dimensional Lattice
Systems and Finite Volume Factorization Properties."
{\sl J.\ Statist.\ Phys.} {\bf 59}, 221--256 (1990).

\bibitem{[S1]}
S.B.\ Shlosman,
``Path large deviation and other typical properties of the
low--temperature models, with applications to the weakly Gibbs states."
{\sl Markov Process.\ Related Fields} {\bf 6}, 121--133 (2000).



\bibitem{[ScS]}
R.H.\ Schonmann, S.B.\ Shlosman,
``Complete analyticity for $2$D Ising completed."
{\sl Comm.\ Math.\ Phys.} {\bf 170}, 453--482 (1995).



\end{thebibliography}
\end{document}